\documentclass[fleqn,10pt]{wlscirep}
\usepackage[utf8]{inputenc}
\usepackage[T1]{fontenc}
\usepackage{graphicx}
\usepackage{hyperref}
\usepackage{subcaption}
\usepackage{jabbrv}

\title{Distributed Acoustic Fiber Sensing for Research Campuses and Large Scientific Infrastructures - The Hamburg WAVE proto-network}

\author[1]{Oliver Bölt}
\author[2,\dag]{Luigia Cristiano}
\author[3]{Sandy Croatto}
\author[1]{Dirk Gajewski}
\author[3]{Erik Genthe}
\author[4,*]{Oliver Gerberding}
\author[1,*]{Céline Hadziioannou}
\author[1]{Conny Hammer}
\author[3]{Markus Hoffmann}
\author[3,5,*]{Katharina-Sophie Isleif}
\author[1]{Antonia Kiel}
\author[2,6]{Charlotte M. Krawczyk}
\author[1]{Regina Maass}
\author[1]{Ingra Barbosa}
\author[3]{Norbert Meyners}
\author[5]{Reinhardt Rading}
\author[3,*]{Holger Schlarb}
\author[7]{Roman Schnabel}
\author[5]{Wanda Vossius}
\author[2]{Christopher Wollin}

\affil[1]{University of Hamburg, Institute of Geophysics, Bundesstrasse 55, Hamburg, 20146,Germany}

\affil[2]{GFZ Helmholtz Centre for Geosciences, Telegrafenberg, Potsdam, 14473,Germany}

\affil[3]{Deutsches Elektronen-Synchrotron (DESY), Notkestraße 85, Hamburg, 22607, Germany}

\affil[4]{University of Hamburg, Institute of Experimental Physics, Luruper Chaussee 149, Hamburg, 22761, Germany}

\affil[5]{Helmut-Schmidt Universität, Institut für Automatisierungstechnik, Holstenhofweg 85, Hamburg, 22043, Germany}

\affil[6]{TU Berlin, Institute for Applied Geosciences, Ernst-Reuter Platz 1, Berlin, 10587, Germany} 

\affil[7]{University of Hamburg, Institute for Quantum Physics, Luruper Chaussee 149, Hamburg, 22761, Germany} 

\affil[$\dag$ ]{now at LIAG Institute for Applied Geophysics, Stilleweg 2,  Hanover, 30655, Germany }

\affil[*]{oliver.gerberding@uni-hamburg.de}
\affil[*]{celine.hadziioannou@uni-hamburg.de}
\affil[*]{isleifk@hsu-hh.de}
\affil[*]{holger.schlarb@desy.de}

\begin{abstract}

We demonstrate and investigate how Distributed Acoustic Sensing (DAS) can be utilized on research campuses and in large scientific infrastructures to study environmental vibrations and reduce their impact on high-precision experiments.

We first discuss the potential of DAS in the context of particle accelerators, gravitational wave detection experiments and research campuses. 

Next, we present the results of our seismic measurement campaign conducted with our proto-network, which involved the probing of over 12 km of fiber, in May 2021. This campaign was conducted by the Hamburg WAVE initiative in Science City Hamburg Bahrenfeld and included DESY, the European XFEL, PETRA III and the University of Hamburg.

Our proto-network confirms the ability to observe natural, anthropogenic, and infrastructural vibrations and how and where these couple into different parts of the heterogeneously set up fiber network. 
We also present results on a study of noise and motion coupling aspects of DAS probing double-redundant fiber loops in a unique environment,
the European XFEL.

Our results show that DAS can greatly benefit research campuses and large scientific infrastructures and they highlight the opportunities and challenges of implementing and operating such seismic networks. 

\end{abstract}

\begin{document}

\flushbottom
\maketitle

\thispagestyle{empty}

\section{Introduction}\label{intro}

As high-precision experiments, instruments and research facilities become ever more sensitive, they also become more susceptible to external and environmental disturbances, especially ground vibrations from seismic waves. Research campuses and large facilities like particle accelerators, which host multiple experiments in close proximity, are particularly affected and can greatly benefit from seismic monitoring of vibrations. These facilities are often located near cities, so they experience both natural and human-made vibrations, including those from their own activities or nearby sources in their surroundings. Ultra-sensitive experiments, like gravitational wave observatories, also benefit from seismic monitoring. Even though these observatories are typically in quieter, remote areas, their sensitivity to disturbances both internal and externally generated, is generally much higher.

Dense seismic networks on research campuses and in scientific infrastructures can detect and mitigate disturbing vibrations. 
The type of sensing network we propose and test here shall combine the ability to identify and locate sources of vibrations with very high spatial and temporal resolution by employing distributed acoustic sensing (DAS), which enables us both to reduce their influence on experiments and to study these sources and the propagation of their induced structural and seismic waves in great detail. 

In Section \ref{sec:motivation}, we briefly review the capabilities and uses of DAS. We then discuss seismic-network studies for gravitational-wave detectors and vibration sensing in particle accelerators. Finally, we explain the motivation for our study by focusing on dense sensing networks on research campuses, especially our own campus in Hamburg. In Section \ref{sec:protnetwork} we describe the set-up for the proto-network seismic measurement campaign we conducted in May of 2021 in the Science City Hamburg Bahrenfeld. The results of our measurement campaign are presented in Section \ref{sec:results}, where we first report on various vibrations measured in and around our facilities and then describe a DAS self-noise characterization along the tunnel of the European XFEL. Section \ref{sec:conclusion} finally concludes with an outlook and a summary.

\section{Motivation}
\label{sec:motivation}

DAS has emerged as a groundbreaking technology for establishing large $N = \mathcal{O}(10^3-10^4)$  vibration sensor networks, leading to significant advancements in various geophysical and seismological experiments \cite{Hartog2017,lindsey2020,lindsey2021,martin2021}. 
DAS devices measure the ground-motion-induced stretching of deployed glass fibers.
State-of-the-art commercial DAS devices measure the phase change of pulsed laser light that is back scattered from intra-fiber scatterers, creating a distributed vibration sensor. 
One promising application of DAS is the utilization of unused telecommunication fiber optic cables, known as ``dark fiber''.
Although not purposefully installed for seismological measurements and therefore subject to site-specific variations in recording quality \cite{AjoFranklin2019}, these systems can span up to several dozen kilometers, enabling large-area investigations with minimal effort and cost.

DAS converts an optic fiber into a dense seismic array with regularly spaced sensors along the fiber thus facilitating the coherent 
and spatially un-aliased recording of the seismic wavefield. This enables imaging of geologic features, like faults \cite{jousset2018, Currenti2021}, low detection thresholds in seismic monitoring \cite{Lellouch2021}, efficient densification of geophone networks in response to a seismic crisis \cite{Flovenz2022}, and more accurate observations of atmospheric phenomena such as meteoroids and thunderstorms \cite{VeraRodriguez2022, Zhu2019}.
Moreover, studies focusing on urban environments have examined the retrieval of signals created by anthropogenic activity \cite{martin2018} and have been applied to seismic site characterization and exploration  \cite{spica2020, dou2017, Cheng2023, Ehsaninezhad2024,10.1785/0220250056}, structural health and traffic monitoring \cite{Liu2023}, as well as environmental sensing in urban experiments \cite{zhu2021,Krawczyk2019}. Notably, ongoing research has explored the implementation of vibration sensor networks in specific settings, such as monitoring machinery through noise interferometry on a research campus \cite{zhuxiao2022}, and investigating ambient seismic noise in an urban environment using downhole DAS at a university campus in Perth, Australia \cite{shulakova2022}. These studies collectively demonstrate the extensive range of applications and the significant impact of DAS in advancing geophysical research, urban monitoring and the potential for future smart research campuses that could sense as well as adapt and react to changes in their working environments.

Vibration sensor networks are also prominently studied and employed in the context of seismic and Newtonian noise suppression in ground-based gravitational wave detectors \cite{abbott2016a,badaracco2019}. Seismic noise reduction benefits from feed-forward signals, especially at low-frequencies where inertial sensors are often self-noise limited \cite{matichard2015b}. Newtonian noise is a force-noise and therefore couples inherently stronger at lower frequencies. Since it cannot be shielded, mitigation strategies rely on  predictions from measurements of the incoming seismic waves \cite{trozzo2022}. Current sensor array implementations, such as the Newtonian noise sensor array in Virgo \cite{badaracco2020}, focus on relatively high vibration frequencies and are, correspondingly, limited in terms of their spatial extent. Larger arrays are studied in the context of future third generation gravitational wave detectors, such as the Einstein Telescope \cite{punturo2010}, where Newtonian noise will be suppressed at frequencies below 10\,Hz, requiring kilometer-scale sensing arrays. 
Sensor arrays are also examined in the broader context of environmental noise at detector sites \cite{fiori2020,nguyen2021} and will be essential for foreseen upgrades of current detectors and the design of future ones, including 4th-generation detectors operating below 1\,Hz \cite{harms2013}. DAS has been proposed as a possible technique to realize or complement such sensor arrays \cite{forbriger2025}.

With the increasing circumference and decreasing beam size of envisaged particle accelerators for high energy physics (HERA 6.3\,km \cite{HERA1981Proposal}, LHC 27\,km \cite{LHC2004DR},
SSC 87.1\,km \cite{SSC1986CDR}) the effect of ground motion on the performance has been studied as early as in the 1980s \cite{rossbach1987closed,LHC2005GManalysis,SSC1990GM}.
In 1990s the proposals to build a 30\,km long straight linear particle collider for high energy physics with very small beam sizes at the collision point in the center (JLC \cite{JLC1997Dstudy}, NLC \cite{NLC2001CDR}, TESLA \cite{TESLA2001TDR}, CLIC \cite{CLIC2012CDR2}, ILC \cite{ILC2013TDR2}) led to intensified investigations of vibration couplings and reduction schemes \cite{SeryiA2001NLC-GM,Bialowons2006MEASUREMENTOG,Snuverink2011CLIC,GMonLC1996}.
Vibrations impact accelerator facilities by inducing misalignment of its components and thereby causing beam-instabilities that, in-turn, can create noise, background signals and put additional stress on the beam control and feedback systems. 
With the invention of so-called multi-bend achromats in the accelerator design to achive very small particle beams for accelerator-based synchrotron light sources \cite{Einfeld2014MultiBendAchromat} the influence of vibrations again moved into the focus \cite{sajaev2018APS-U-GM,SSRF2019GMeffects}. 
The research center DESY~\cite{habfast2013grossforschung}, which is situated west of Hamburg, Germany, is the core of the research campus Science City Hamburg Bahrenfeld and hosts a number of particle accelerators, such as the free-electron lasers European XFEL \cite{altarelli2007european,decking2020mhz} and FLASH \cite{tiedtke2009soft} and the synchrotron light sources PETRA \cite{balewski2004petra} and, in the future, PETRA IV \cite{Schroer:426140}.
The use of a vibration sensor network based on distributed acoustic fiber sensors has 
not yet been studied extensively
in the context of such accelerators, but can provide major benefits to monitor the accelerator infrastructure and to study the coupling of vibrations into the machines. We already demonstrated that DAS measurements can be used to identify ocean generated microseism as a major contribution to the arrival time jitter of the EuXFEL \cite{Genthe_Czwalinna_Lautenschlager_Schlarb_Hadziioannou_Gerberding_Isleif_2025}.

Research campuses offer a unique environment for seismic sensing, combining high-precision requirements of measurements with the potential for dense sensor deployments in long-term implementations. Such configurations can support sustainable and optimized experimental designs and conditions. Moreover, campuses often have an abundance of unused telecommunication fiber that could be repurposed as seismic sensor networks. 
At the same time, the heterogeneous topological markup of campuses, which is often shaped by the historical growth of buildings and infrastructure, pose significant challenges for studying wave propagation. These challenges can be addressed through sufficiently extensive and broad sensor coverage, as we envision. The authors in Fang et al. (2020) \cite{fang2020} demonstrated the successful application of DAS for urban near-surface seismic monitoring by re-purposing existing telecommunication fiber-optic cables beneath Stanford University. Their work achieved high-resolution subsurface imaging through ambient noise interferometry and revealed seasonal velocity variations linked to groundwater dynamics. Their work also highlights DAS's potential to transform existing urban infrastructure into scalable seismic networks for smart-city hazard mitigation and environmental monitoring. We examine whether this capability can be applied directly to science campuses, where it could also enhance the performance of local experiments and research facilities. However, the diverse spectrum of experiments on each campus requires an interdisciplinary approach to designing, deploying, and using such sensor networks. As for the Science City Hamburg Bahrenfeld this includes, among others, accelerator physics, photon science, gravitational wave detection physics, astro-particle experiments, geophysics and engineering. Some public results of seismic measurements from the research campus and the Hamburg WAVE initiative can be found at: \href{http://www.wave-hamburg.eu/}{\bfseries wave-hamburg.eu} .

\begin{figure}[h!]%
\centering
\includegraphics[width=0.99\columnwidth]{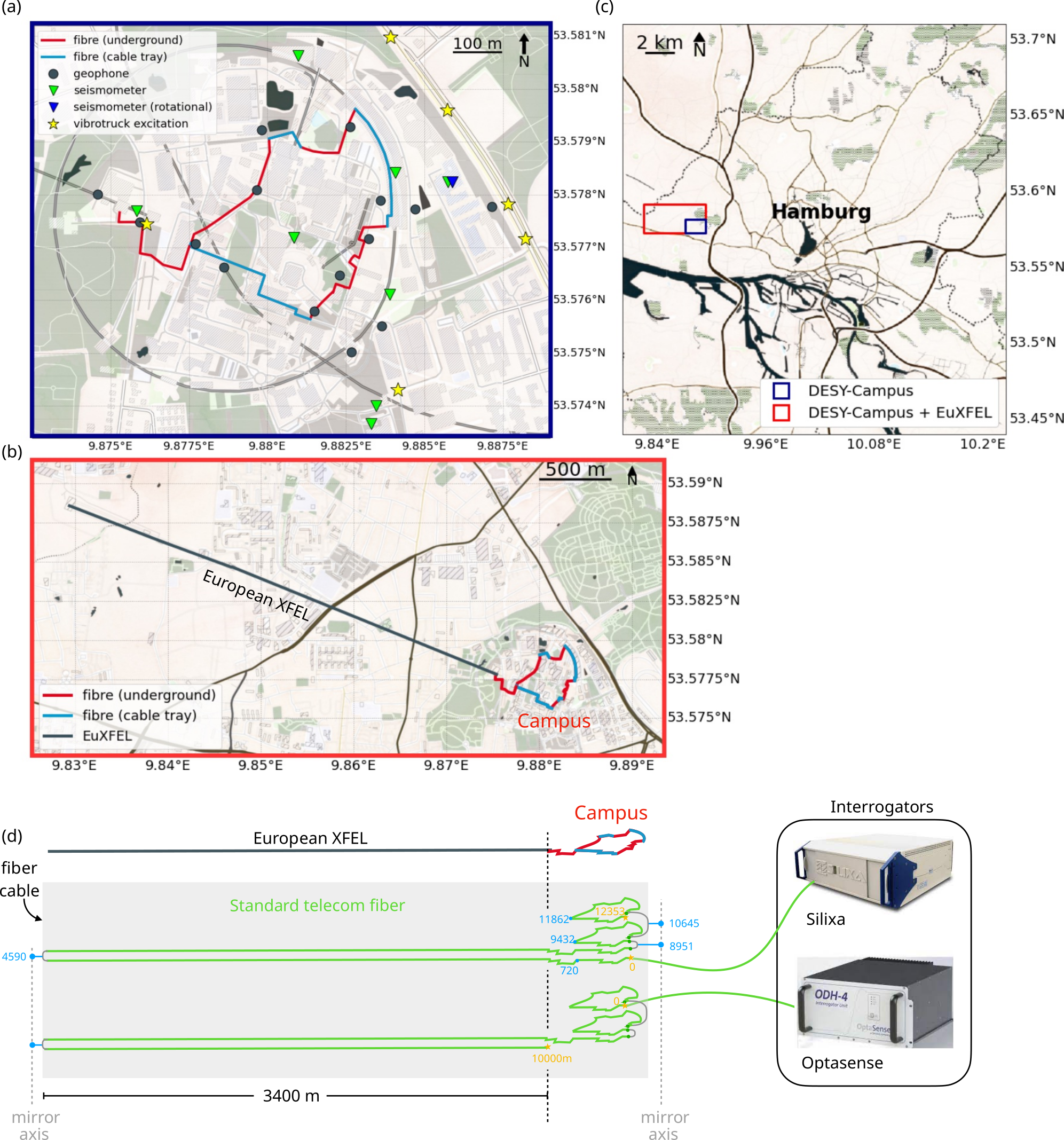}
\caption{Map of the fiber route during the demonstration study in the Science City Hamburg Bahrenfeld, on the campus of DESY and University of Hamburg, and in the tunnel of the European XFEL. In the north-east the fiber is routed through the experimental hall of the PETRA III accelerator, where the beam lines are situated, and follows the circular shape of the ring. The additional sensors and vibro-truck excitation points are also marked, as is the routing of the fiber either underground in ducts or over ground in cable trays. 
The maps (a) and (b) show details of the campus and a wider view of the surrounding area including the XFEL tunnel. Map (c) shows the location of (a) and (b) within the metropolitan area of Hamburg, which is roughly 100\,km away from the North and the Baltic sea.
\newline The lower sketch (d) illustrates the fiber paths. Multiple fibers within a single fiber optic cable were spliced together to form one loop. Another set of fibers, also spliced together within the same optical cable, was monitored using a second interrogator system.}\label{fig:map}
\end{figure}

\section{The proto-network seismic measurement campaign}
\label{sec:protnetwork}

Here we report on a two-week demonstration study that was conducted in May 2021 in the Science City Hamburg Bahrenfeld. The measurement campaign included distributed acoustic sensing of 12.132\,km of fiber, which formed redundant loops, with simultaneous use of two types of DAS interrogators. We also deployed 15 Geophones, 11 broadband seismometers and one rotational seismometer on the campus. During the deployment the network recorded six active vibro-truck excitations at different spots on the campus, as well as one larger earthquake (China, 2021-05-21, 18:00 - 19:30 Central European Time, magnitude 7.4). The fiber was routed through ducts, on cable trays outdoors and inside the PETRA III experimental hall, and both below the tunnel floor and on cable trays in the European XFEL tunnel (3.4\,km physical length), as illustrated in Figure \ref{fig:map}.

The backbone of our distributed sensor network comprised both dedicated and repurposed telecommunications fibers.
Each fiber segment shown consisted of four fibers, two of which we spliced together such that we created two fiber loops that redundantly probed each segment in forward and backward direction. Each loop was connected to one of the two deployed DAS interrogators, Silixa MK1 iDAS \cite{silixa_www} and Optasense ODH4 \cite{optasense_www}, such that almost all fiber segments shown in the map were effectively interrogated four times (one of the interrogators could not probe the full loop length during the campaign, hence some part of the fiber route was only sensed three times). 
Note that the spatial distance between channels is different from the optical distance traveled by the probing light pulse from the interrogator to the respective channel.
With the Silixa MK1 iDAS interrogator we continuously recorded data along a 12.132\,km long fiber between May 17th -- 31st 2021, with a sampling rate of 1000\,Hz. 
The iDAS uses a gauge length of 10\,m and the spatial sampling was set to a channel spacing of 1\,m.
We used the Optasense ODH4 for continuous data acquistion with a sampling rate of 10000\,Hz along a 10\,km fiber sub-segment for one week between May, 24th -- 28th, 2021, with the same spatial sampling parameters. 
The 15 3-component, 4.5~Hz Geophones were placed on soil throughout the campus, the 11 broadband seismometers of different types were placed in buildings on campus. One broadband seismometer, a Trillium Compact 120S, was collocated with a rotational seismometer, iXblue blueSeis-3A, in a laboratory building for gravitational wave detection in the north-east, to provide six-degree of freedom ground-motion measurements. All Geophones and seismometers were deployed and recorded data for the whole two-week period in which the Silixa iDAS V.2 interrogator was operating. 
These point sensors, recording ground velocity or acceleration, complement the dense strain-rate measurements of DAS. The relation between both quantities and dedicated comparisons are discussed in \cite{wang2018ground,lindsey2020broadband,kennett2024guide,rading2025distributedacousticsensingenvironmental,vossius2026distributed}.

\begin{figure}[t]
    \centering
    \begin{minipage}{0.7\textwidth}
        \includegraphics[width=\linewidth]{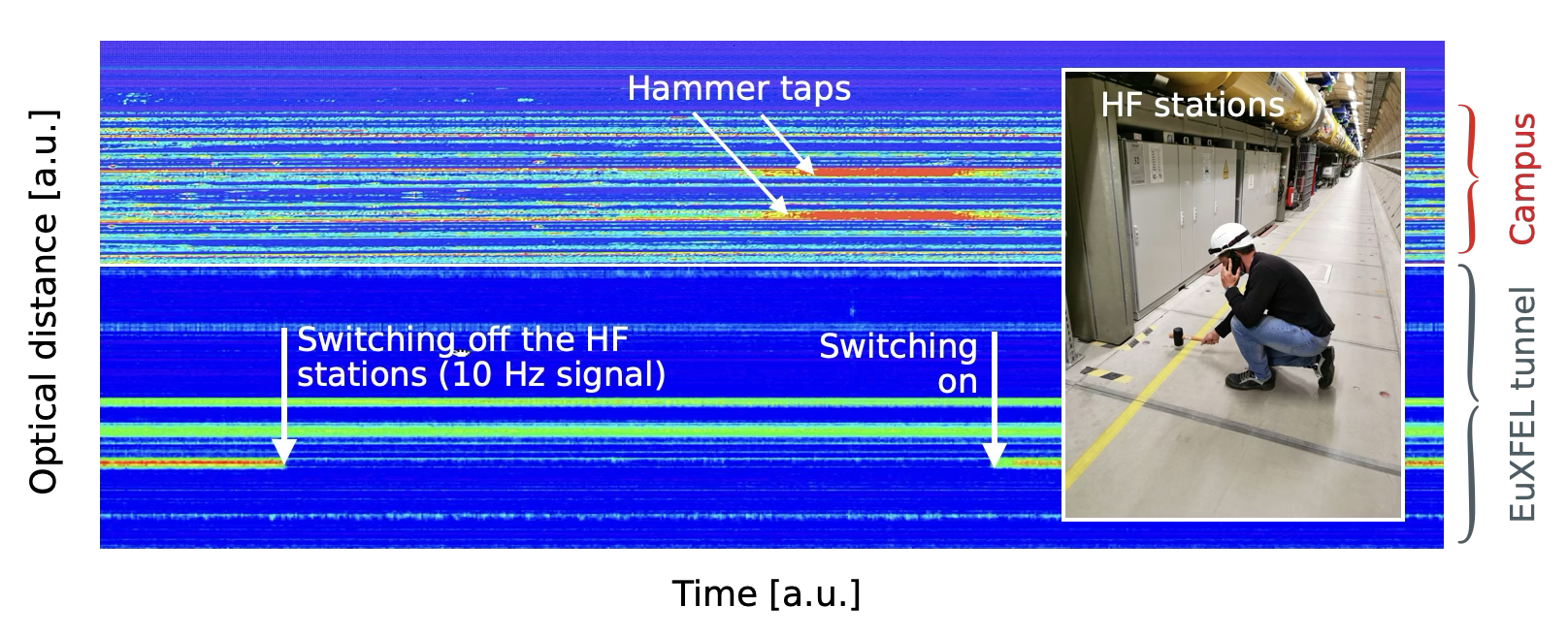}
    \end{minipage}%
    \hfill
    \begin{minipage}{0.27\textwidth}
        \captionof{figure}{Screenshot of the waterfall plot showing geo-referencing of the DAS fiber loop using hammer taps on campus and research infrastructure, such as HF stations in the EuXFEL, where the hammer taps were not visible due to the concrete floor. }
        \label{fig:geo}
    \end{minipage}
\end{figure}

After setting up the DAS interrogators we used GPS-localized hammer blows along the fiber path to geo-reference the interrogated fiber segments to map locations as shown in Figure \ref{fig:geo}. Inside the European XFEL tunnel, where GPS was unavailable, geo-referencing was performed relative to the tunnel’s coordinate system using hammer impacts. The geo-referencing within the tunnel was later also verified by the localization of specific vibration sources along the tunnel which could be associated to specific parts of the accelerator.

During the campaign about 60 Terabyte of raw data were acquired. Live data analysis largely consisted of monitoring waterfall plots of RMS vibration amplitudes along the fiber path that are conveniently computed and projected in real-time by both DAS systems. 

In addition to observing the occurring natural, anthropogenic and infrastructure vibrations we conducted a number of smaller experiments, such as intentional car rides on a street above a fiber segment, additional hammer blows and the activation and deactivation of high-frequency stations of the European XFEL. We also witnessed two shutdown periods of the European XFEL and six vibro-truck excitations, four of which took place along one of the main roads near the campus and one was conducted close to the European XFEL tunnel. 

\section{Results}
\label{sec:results}

\subsection{Vibrations in and around large scale research facilities and our campus}

\begin{figure}[h!]%
\centering
\includegraphics[width=\columnwidth]{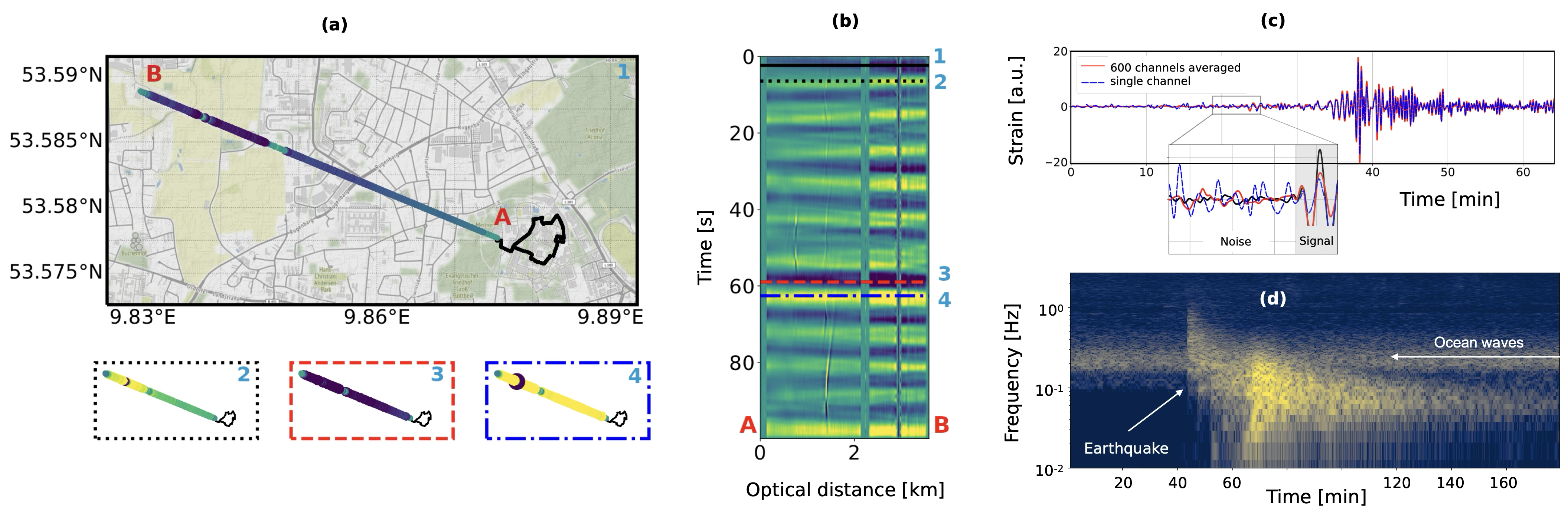}
\caption{Spatiotemporal propagation of seismic waves through the European XFEL tunnel (a) caused by a strong earthquake (magnitude 7.4) with its epicenter in Qinghai, China. A typical representation for DAS data (waterfall diagram) is shown in (b), which color-codes the time series of strain amplitude (y-axis) for each fiber sensor (channel, x-axis). Here, the dark blue color represents compression and light yellow represents dilatation of a fiber segment. Stationary sources such as pumps thus appear as stripes at a fixed position, while moving sources and propagating waves appear as inclined lines whose slope gives their apparent velocity.
Four different points in time, marked by numbered horizontal lines in the waterfall diagram, correspond to snapshots of strain rate amplitudes of all DAS sensors in (a).
These are plotted in the same colormap, projected to their corresponding location above the fiber trajectory in the European XFEL tunnel. Here, the colors and dashes of the horizontal lines and frames of the snapshots match. Although the wavelength exceeds that of the tunnel considerably, the spatial shaping of the wave crests and troughs can be seen at a fixed point in time (animation available at \href{http://www.wave-hamburg.eu/}{\bfseries wave-hamburg.eu}). \\
(c) shows an illustration of SNR enhancement through stacking of 600 adjacent DAS channels, compared to single channel waveform. (d) shows a spectrogram of single DAS channel recording, showing both P- and S-wave arrivals of the earthquake, along with the permanently active ocean microseism band. 
}\label{fig:eq_snapshots}
\end{figure}

We showcase the DAS installation's practical capabilities during the magnitude 7.4 Qinghai earthquake on the 21st of May 2021, shown in Fig.~\ref{fig:eq_snapshots}.
Our DAS measurements recorded high-resolution strain data that revealed clear seismic wave propagation patterns along the European XFEL. The dense spatial sampling of DAS along the fiber section enabled the observation of the direction of incidence and propagation from east to west of the seismic waves coming from China.
The independently determined hypocenter provided a reliable reference point for verifying DAS performance. This event underscores the potential of DAS for earthquake early warning systems \cite{10.1785/0120210214}, offering dense spatial coverage and the ability to significantly enhance the signal-to-noise ratio through channel stacking \cite{diaz-meza2023, maass2024}, as illustrated in Fig.~\ref{fig:eq_snapshots}, which compares the signal from a single DAS channel to the combined signal of 600 stacked channels.
Furthermore DAS is most sensitive to axial strain, making it particularly responsive to P-waves when the fiber is aligned parallel to the direction of wave propagation. Sensitivity to S-waves, in contrast, depends on the polarization angle relative to the cable orientation \cite{li2022distributed, hubbard2022}. Nevertheless, the spectrogram in Fig.~\ref{fig:eq_snapshots} clearly shows the system’s sensitivity to both P- and S-waves generated by the earthquake.
The simultaneous detection of both P- and S-waves is particularly valuable, as it enables more accurate earthquake characterization, including precise event localization and wavefield reconstruction. In the context of seismic hazard these capabilities enable improved early warning, which can be utilized on a smart-campus to feed improved control loops to stabilize measurements and machines.
The spectrogram in Fig.~\ref{fig:eq_snapshots} reveals another persistent natural noise source, microseisms generated by ocean waves \cite{ardhuin2015,juretzek2016}, which appear continuously in the data with varying intensity. This continuous, low frequency signal was identified as a limiting factor in the EuXFEL femtosecond synchronization. With the help of DAS it was shown that ocean-generated microseism has to be compensated to achieve the 1 femtosecond goal in bunch arrival stability at the EuXFEl \cite{Genthe_Czwalinna_Lautenschlager_Schlarb_Hadziioannou_Gerberding_Isleif_2025}.

\begin{figure}
    \centering
    \includegraphics[width=\linewidth]{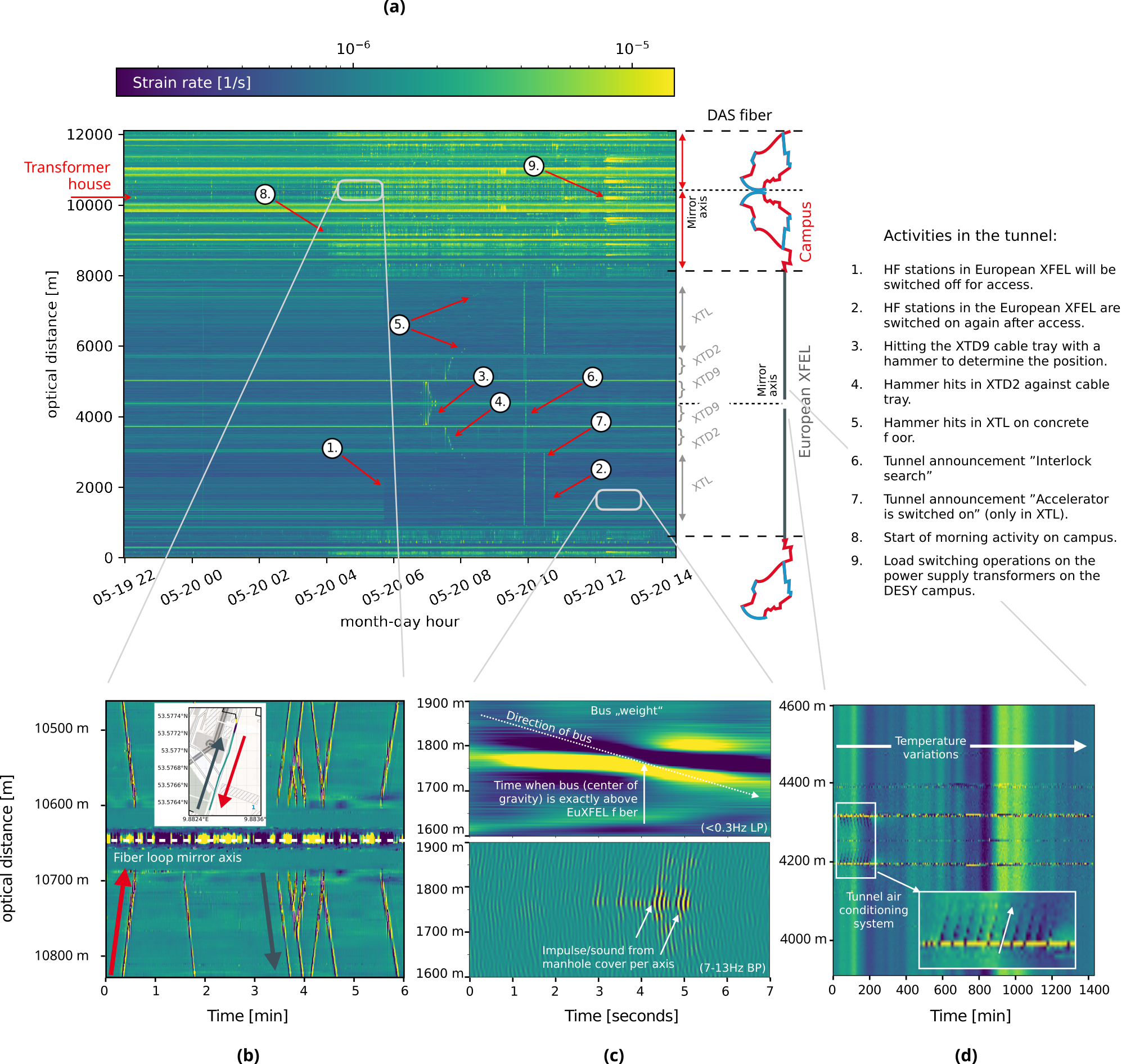}
    \caption{Waterfall plots of the strain rate measured by the DAS network along the 12.132\,km-long fiber. Several activities—highlighted in (a) are directly visible without the need for extensive data processing. Additional events are shown in (b)–(d), where low-pass (LP) filters were applied to enhance the visibility of traffic-induced vibrations and temperature-related variations. (a) Overview of noise and signal as waterfall plot of all DAS sensors/channels over the 12.132km fiber. Max. amplitude strain rate of 20s intervals is color coded. (b) Passing cars close to the mirror axis with a speed of $18\pm 5\,$km/h. (c) Passing bus on Flurstraße 10m above the EuXFEL. (d) Longterm (1 minute) averaged strain rate variations along the XTL tunnel showing temperature changes.
    The rectangles indicating the location of the sub-figures (b) and (c) in the main plot (a) are not in scale.}
    \label{fig:overview_time}
\end{figure}

\begin{figure}
    \centering
    \includegraphics[width=\linewidth]{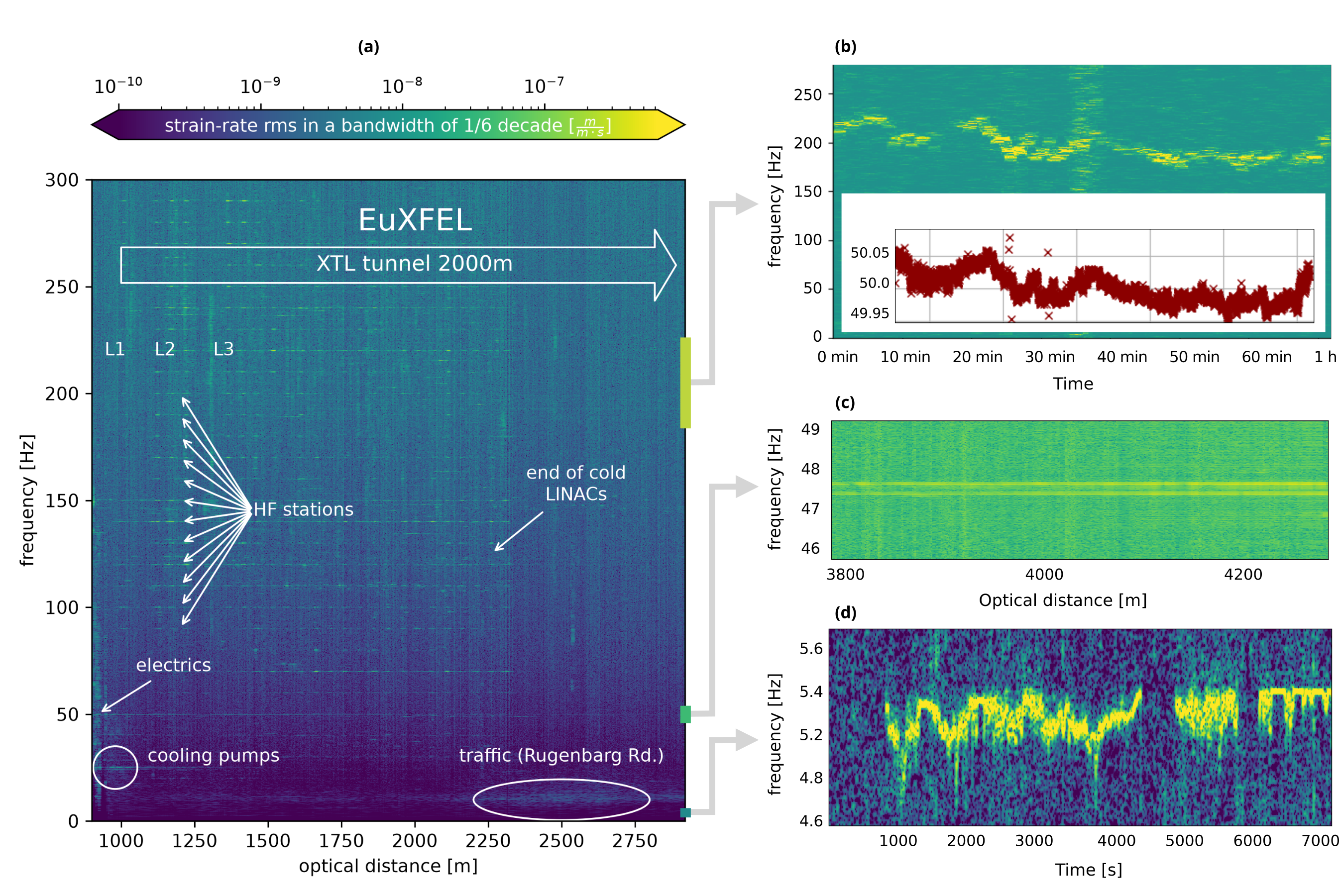}
    \caption{Spectrograms for parts of the DAS network. Frequency dependent amplitudes varying over distance in the fiber are shown in (a) and (c). Variations of the frequency dependent amplitude over time for specific points in the fiber are shown in (b) and (d) to visualize temporal variations of signals. (a) DAS spectrogram along the EuXFEL. (b) DAS spectrogram over time shows the 200 Hz harmonic of the 50 Hz power line signal measured near a tranformer house.   
   The inset shows the transformer house's mains 50\,Hz fluctuations measured in the DESY power supply system. (c) 47.5\,Hz oscillation caused by Helium compressors in the tunnel. (d) Snapshot of 5.2\,Hz oscillation disturbing PETRA III operation.}
    \label{fig:spectrogram}
\end{figure}

Two vibration maps, one in the time–space domain presented as a waterfall plot shown in Fig.~\ref{fig:overview_time}~(a) and one in the frequency–space domain shown in Fig.~\ref{fig:spectrogram}~(a), reveal numerous noise and vibration sources across the research campus, even without the application of advanced data processing or filtering algorithms. Fig.~\ref{fig:overview_time}~(a) displays 16 hours of data recorded along the entire 12.132\,km-long fiber, showing only the maximum amplitude within each 20\,second interval at the corresponding measurement point. Continuous vibrations, such as those produced by technical infrastructure, can be directly identified based on signal strength and spatial location. Data from the DAS system has the potential to provide a comprehensive overview of the activities taking place in the accelerator tunnel, especially when presented in the frequency domain across different channels. Such visualizations offer a rich and expressive representation of the actual events unfolding within the tunnel environment as shown in Fig.~\ref{fig:spectrogram}~(a).

Fig.~\ref{fig:overview_time}~(a) reveals a clear distinction between day and night activity on the campus, with an increase in noise levels during morning hours as daily operations begin. In addition, components of the research infrastructure, such as the high-frequency (HF) stations of the European XFEL, can be monitored using DAS. Their switching on and off is visible both in Fig.~\ref{fig:overview_time}~(a) at positions (1) and (2), and in Fig.~\ref{fig:spectrogram}~(a), where higher harmonics of the HF stations are marked. These stations are located exclusively in the so-called cold LINAC sections, whose spatial extent is identifiable in the spectrogram. Tunnel announcements are also detectable using DAS, as shown in Fig.~\ref{fig:overview_time}~(a) at positions (6) and (7), indicating the specific regions of the tunnel where announcements were made, once in the entire European XFEL facility (6), and once limited to the XTL section (7).

 The detection of vehicles using DAS is a known capability\cite{dou2017} and can be measured and spatially resolved in our set up. Fig.~\ref{fig:overview_time}~(b) displays vibrations caused by vehicles passing on a road section directly adjacent to the fiber-optic cable. This demonstrates the ability of DAS to localize moving vibration sources with high precision, enabled by the dense spatial sampling of the sensor array. 
Notably, the recordings enable the differentiation of vehicle types, namely cars, bicycles and busses.  
 One example, shown in Fig.~\ref{fig:overview_time}~(c), captures a bus, characterized by its two axles, crossing a manhole cover on a road located approximately 10 meters directly above the European XFEL tunnel. The resulting impulses due to the two axles crossing is visible as two signals in the lower plot, bandpass filtered between 7\,Hz and 13\,Hz, while the upper plot (low-pass filtered below 0.3\,Hz) illustrates how the weight of the bus induces slow strain on the fiber and the concrete tunnel itself.

Fig.~\ref{fig:overview_time}~(d) shows long-term temperature variations and the operation of the tunnel’s air conditioning system, revealed by temporally averaging the strain rate measurements over one minute along the XTL tunnel section of the European XFEL. This approach allows tracking the spatial distribution and movement of conditioned air within the tunnel, as highlighted in the figure.

Multiple signals can be distinguished in the frequency-space spectrogram, as shown in Fig.~\ref{fig:spectrogram}~(a) for parts of the European XFEL. Continuous and random vibration sources, such as cooling pumps and road traffic along the nearby Rugenbarg street, can be identified.
This enabled us to identify and study also unexpected signals, as discussed for three examples in the following.

A transformer house located on campus is positioned directly above a segment of the fiber-optic cable. This electrical system operates at alternating voltages with a mains frequency of around 50\,Hz, as shown in the inset of Fig.~\ref{fig:spectrogram}~(b); however, the non-monochromatic nature of the resulting interference gives rise to higher-order harmonics of the fundamental frequency, which can extend into the kilohertz range\cite{cohen2010} and cause corresponding mechanical vibrations. In the DAS recordings the 4th higher harmonic of 200\,Hz is shown as an example in Fig.~\ref{fig:spectrogram}~(b). Temporal variations in the electronically measured 50\,Hz signal show strong correlation with the DAS response. This availability of continuous seismic data recorded at a high spatial resolution on the campus would enable coherence analysis with electrical signals, facilitating the design of targeted filters \cite{cohen2010} and acoustic insulation to suppress noise and enhance the quality of high precision experimental measurements. 

Experimenters at the European XFEL reported interference signals that appeared as artefacts in their experimental data, impairing measurement quality. One prominent interference was a persistent hum at 47.5\,Hz, observed in the experimental hall at the end of the accelerator, with initially unknown origin. This acoustic noise was evidently transferred to the photon beam and ultimately manifested as artefacts at the experimental stations.
As shown in Fig.~\ref{fig:spectrogram}~(c), this interference frequency is also visible in the DAS data. The spatial resolution of DAS enabled detailed examination of both the nature of the disturbance, seismo-acoustic waves, and their propagation along the tunnel. Notably, the interference signal was detectable over several hundred meters, highlighting the efficiency of vibration transmission in the tunnel environment. Analyzing the amplitude decay of the vibration provided an indication for the location of the source. Subsequent investigations at European XFEL identified helium compressors as the source of the 47.5\,Hz oscillation. While DAS was not the exclusive tool used for source localization in this case, it proved highly valuable for detecting, characterizing, and understanding unknown noise sources in complex experimental infrastructures.

At the PETRA experimental stations interrogated at a spatial distance in our fiber of approximately 10 km (Fig.~\ref{fig:overview_time}~(a)), a persistent interference signal with a frequency around 5.2\,Hz was recorded. The temporal evolution over approximately one hour is shown in Fig.~\ref{fig:spectrogram}~(d). Although this signal disturbed measurements at PETRA its (seismic) origin remained unclear until identified by the experiment outlined here. 
Over rather long distances > 3\,km this signal appeared over extended periods on multiple days, with irregular interruptions, and posed a significant challenge for experimental operations of the accelerator, especially due to its temporal variability. The DAS data enabled clear detection and characterization of this interference and revealed its seismic nature. Detailed DAS measurements from the here reported campaign indicated that the disturbance propagates as a ground vibration. Source of the disturbance was a vintage helium compressor from 1965, situated on campus, that was manually and only occasionally operated. The location of the compressor could not yet be determined with the here reported fiber optic network because of the essentially one-dimensional geometry and the inhomogeneous coupling of the fiber to the ground. A larger spatial fiber network with more homogeneous coupling allowed to localize this disturbance later.

\subsection{DAS self-noise characterization along the European XFEL}
The geometrical setup of the network enables characterization of DAS self-noise, particularly in sections 
where we have two or more available fibers within a single cable,
as shown in Fig.~\ref{fig:map}. This results in several co-located fiber segments with reversed readout directions and different optical distances from the interrogator.

\begin{figure}[h!]%
\centering
\includegraphics[width=\textwidth]{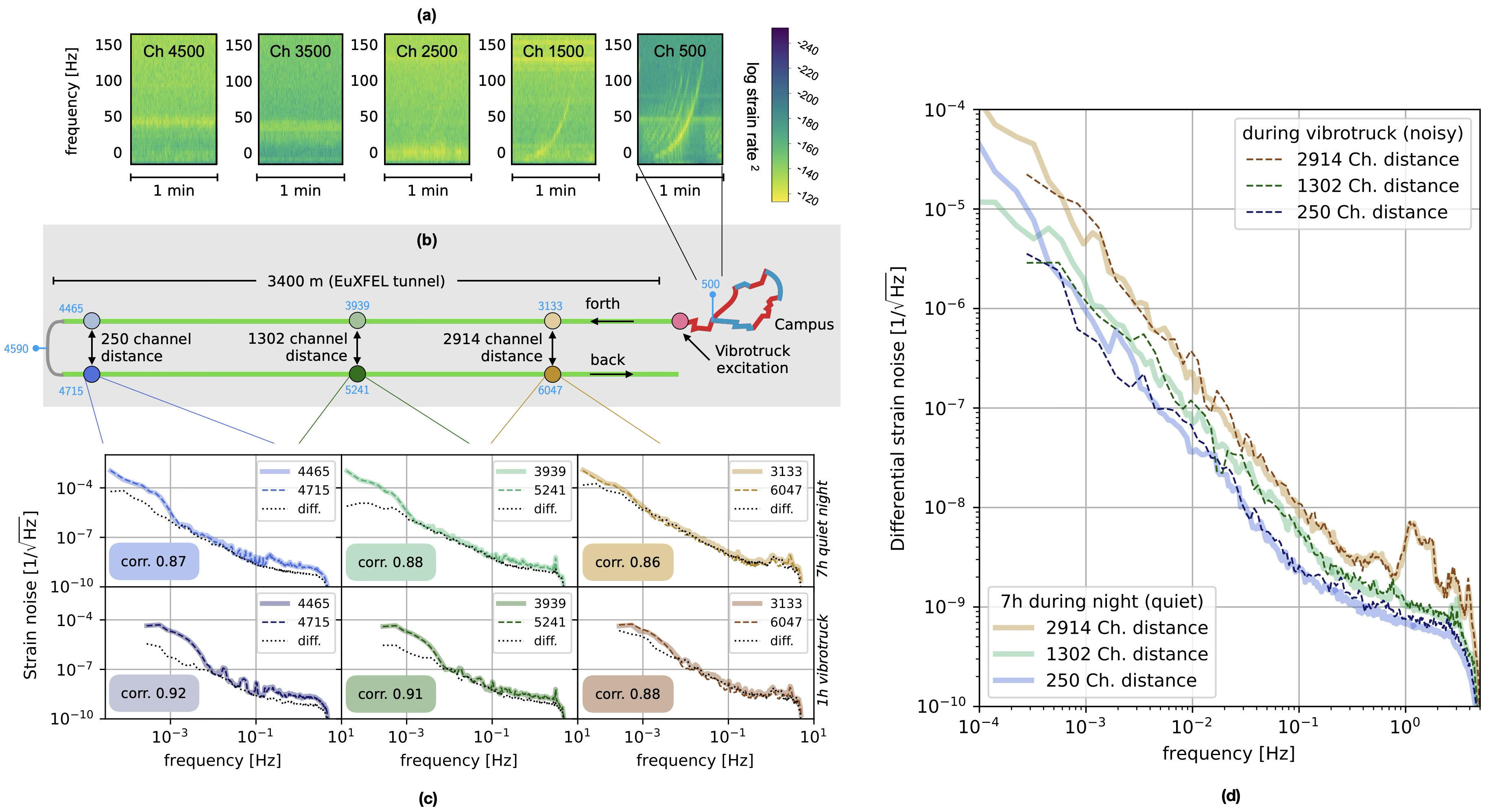}
\caption{Analysis of Linear Power Spectral Densities (LPSD) across fiber channels.
Shown is the LPSD of strain (1\,m baseline) measured at various positions along the fiber. To assess strain noise, two highly correlated fiber segments were selected and their signals subtracted, isolating the differential strain noise. Channel pairs were chosen based on high correlation values, indicating their spatial proximity. The comparison illustrates how differential noise depends on the optical distance between co-located channels, providing insights into the self-noise behavior of the DAS system. (a) Spectrograms during the time of the Vibrotruck sweep along the EuXFEL. (b) Selected DAS channels for data analysis and vibrotruck position. (c) Power spectral densities of the strain rate at two co-located (but separated with respect to the optical distance) channels for a quiet 7h over night measurement (top plots) and 1h vibrotruck excitation (bottom plots). Differential strain rate is plotted in black. (d) Power spectral densities of the differential strain rate between two spatially co-located DAS channels. The residual noise, obtained by subtracting the signals (see (c)), provides an upper bound on the DAS system’s self-noise. The channels distances given in the legend indicate the optical distance between co-located channels, as shown in (b).}\label{fig:lpsd}
\end{figure}

Here, we investigate DAS self-noise within the European XFEL tunnel under both quiet and noisy conditions to assess signal dependencies and potential nonlinearities. For the quiet case, we selected a 7-hour overnight recording with minimal anthropogenic activity. The noisy environment was characterized by an active excitation using a Vibrotruck operated by Baudynamik Heiland \& Mistler GmbH, which generated a controlled frequency sweep from 3\,Hz to 110\,Hz. The resulting signals were recorded along the tunnel and are shown in the spectrograms in Fig.~\ref{fig:lpsd}~(a). 
With increasing distance to the source, higher frequencies are damped stronger, than lower frequencies.
This behavior reflects frequency-dependent propagation of seismic waves in Earth acting like a low-pass filter \cite{gamal2023}.

To estimate DAS self-noise, channels with the highest correlation are selected.
To isolate self-noise as shown in Fig.~\ref{fig:lpsd}~(b), common-mode signals are canceled out by subtraction.
Given the DAS system’s spatial sampling of 1\,m and a gauge length of 10\,m, the maximum spatial offset between co-located channels is 0.5\,m, an offset small enough not to affect self-noise analysis at frequencies below 10\,Hz. This was verified by temporally interpolating the data and re-evaluating both the maximum correlation and the effectiveness of coherent noise subtraction.
 
Fig.~\ref{fig:lpsd}~(c) shows the Linear Power Spectral Density (LPSD) computed from raw strain rate measurements for two co-located DAS channels, along with their differential strain rate (labeled “diff.”) for two scenarios: a 7-hour quiet nighttime period (top) and a 1-hour period during Vibrotruck excitation (bottom). The black dotted line represents the differential strain rate between the two selected channels and is re-plotted in Fig.~\ref{fig:lpsd}~(d) to allow clearer comparison across both cases.
Notably, the Vibrotruck is not the sole source of excitation during daytime measurements. This is evident from increased correlation values observed in all data series during day, even for channels located far from the Vibrotruck. For instance, channels 4465 and 4715 are situated near one of the European XFEL laboratory halls, where regular daily activity likely contributes to coherent signals or anthropogenic noise, increasing the overall correlation between co-located channels.
Fig.~\ref{fig:lpsd}~(c) shows that common-mode rejection is most effective at distinct spectral peaks and in the very low-frequency range. However, at mid-range frequencies, where no dominant spectral peak is present, the differential spectrum closely follows the raw strain rate signals with a $1/f$ slope. This indicates limited common-mode rejection in this range, likely due to residual incoherent noise between the two co-located fiber segments or the DAS noise. Fig.~\ref{fig:lpsd}~(d) shows that this $1/f$ noise floor increases linearly with channel separation, exhibiting a tenfold increase over a distance from 250\,m to 2914\,m. This behavior suggests that the excess noise may originate from sources such as fiber length noise, laser frequency noise, phase readout fluctuations or electronic in the interrogator system. As such, these measurements provide an upper bound on the DAS system’s self-noise. Alternatively, the noise could result from different transfer functions between the fibers and the ground; however, this effect is expected to be minimal, as both fibers are embedded in the same optical cable and should therefore experience nearly identical coupling conditions, which should also be independent of channel separation.
Furthermore, the residual noise observed after common-mode rejection is similar across both measurement conditions (quiet nighttime and Vibrotruck excitation) suggesting a predominantly linear response of the DAS system over the frequency range considered in this analysis.
However, the limited common-mode rejection below 0.1\,Hz — only about a factor of 10 for the strongest peaks during Vibrotruck excitation — restricts the achievable SNR in DAS measurements and limits their usefulness for correcting experiments affected at those frequencies. Of course channel averaging for long wavelength signals can still lead to much better SNR, as shown for the microseism previously in \cite{Genthe_Czwalinna_Lautenschlager_Schlarb_Hadziioannou_Gerberding_Isleif_2025}. The SNR limitation also affects the effectiveness of channel stacking, as co-located channels measuring the same signal may differ by up to 10\%, reducing the potential SNR improvement. Further investigation is required to better understand and mitigate this residual noise in co-located channels.

\section{Conclusion}
\label{sec:conclusion}

The results obtained with our WAVE proto-network demonstrate its ability to identify and study vibration sources with high temporal and spatial resolution in various environments. 
Such networks offer valuable support for future high-precision experiments by providing detailed vibration monitoring of the environment, enabling both the localization and mitigation of sources that degrade the quality of these high-precision measurements.

Looking ahead, developing more complex and permanent networks with broader spatial coverage will further improve the identification of diverse vibration sources. 
This will improve the resolution of wave propagation and interference effects, which are particularly important given the heterogeneous building and infrastructure topologies typically found on research campuses. As a proof of concept, this study also showed that the DAS self-noise and the fiber-to-ground coupling, particularly at low frequencies, require further investigation to fully exploit the capabilities of such networks.

Immediate advances for such networks include low-latency methods for vibration source identification, capable of providing real-time source behavior and location, when the source can be geo-referenced within the network. When combined with a permanent vibration monitoring, these methods can enable the creation of \textit{vibration weather maps} that combine predictions with rapid analysis. Such tools would empower scientists on campus to plan their experiments more effectively and to compare data with the vibration environment at the time. 
Notorious vibration sources that lead to undesired coupling in e.g. on-campus experiments can be actively mitigated through passive or active isolation techniques. 

In the mid-term the identification of less stationary and more obfuscated, vibration sources on campus and in the infrastructures will become possible, including identification and localization of outside sources. Anomalous urban activity could be flagged through
anomaly detection algorithms and known urban events can be characterized (e.g. cars) through event classification. The effectiveness of such detections will depend on the spatial distribution, self-noise and number of sensors. Practical limits will be set by the computational capacity and the algorithms developed for this purpose.  

Since the requirements of experiments to reduce the influence of vibrations increase, these noise reduction techniques will play a crucial role in experiment isolation strategies on research campuses and in the scientific infrastructures, helping to create low-noise environments even in urban areas or ultra-low noise environments in remote sites of e.g. gravitational wave detectors. This enhances the sustainability and robustness of sensitive experiments and the campuses themselves and can enable to effectively create experimental sites on earth with unprecedented low levels of environmental noise. 

As an added benefit, natural and anthropogenic noise effectively illuminates the subsurface, and this illumination can be leveraged to investigate subsurface conditions and their temporal variations. As the network’s sensitivity and size increases, its capability to monitor both the urban environment and the subsurface would be further enhanced, supporting studies of climate adaptation efforts on the campus, the infrastructures and in the surrounding cities.

Our proto-network experiment demonstrates the potential of a seismic network using DAS for research campuses and large scale research infrastructures to create smart campuses and smart infrastructures, especially relevant for, but not limited to, particle accelerators and gravitational wave detection experiments. This study has motivated the establishment of a permanent and growing seismic sensing network, which is in operation in the Science City Hamburg Bahrenfeld since 2023 (see \href{http://www.wave-hamburg.eu/}{\bfseries wave-hamburg.eu}). We expect that such networks may emerge as state-of-the-art for international research centers, offering various exciting research opportunities with regards to their realization, deployment and exploitation.

\bibliography{bibliography}

\section*{Acknowledgements}

The authors acknowledge the additional vibro-truck excitations provided by the company Baudynamik and Dr.-Ing. Michael Mistler during their unrelated measurement campaign.  This experiment was supported by GFZ Helmholtz Centre for Geosciences with instruments provided by the Geophysical Instrument Pool Potsdam (GIPP). We acknowledge support by the DESY IT group, specifically by Tobias Ladwig, for providing the fiber connections and the experiment space. We further acknowledge the operations team of the European XFEL for supporting the geo-referencing and providing access to the accelerator. We thank Joachim Buelow for helping with the preparation and the set-up of the seismometers. We thank Dr. Martin Karrenbach for useful discussions and support during the campaign. We thank the Physnet computing cluster of Universität Hamburg and the Maxwell computing cluster of DESY for providing computing, data storage and support in setting up the processing architecture. We thank the editor and two anonymous reviewers for comments that helped improve the manuscript. 

\section*{Funding}

This research was supported by the Deutsche Forschungsgemeinschaft (DFG, German Research Foundation) under Germany's Excellence Strategy via the Ideas and Venture Fund and ---EXC 2121 ``Quantum Universe''---390833306, and benefited from funding provided by the European Union’s Horizon research and innovation programme under the Marie Sklodowska-Curie grant agreement No. 955515 (SPIN-ITN, \href{https://spin-itn.eu}{https://spin-itn.eu}). The MIN Faculty of Universität Hamburg through the Dean Prof. Dr. Heinrich Graener funded the use of the OptaSense interrogator.

\section*{Author contributions statement}
D.G., O.G., C.Had., H.S. and R.S. conceived the concept. S.C., D.G., O.G., C.Had., K.-S.I., C.M.K., N.M., H.S., R.S. and C.W. planned and prepared the measurement campaign. L.C., S.C., E.G., O.G., C.Had., K.-S.I., R.M., N.M., H.S. and C.W. conducted the measurement campaign. O.B., S.C., E.G., M.H., A.K., C.M.K., R.M., I.B., R.R., W.V. and C.W. performed the data analysis. D.G., O.G., C.Had., M.H., K.-S.I., C.M.K., H.S. and C.W. supervised the campaign and the data analysis. O.B., O.G., C.Had., C.Ham., M.H., K.-S.I., A.K., R.M., I.B., N.M., R.R., W.V. and C.W. wrote the manuscript and prepared the figures. All authors reviewed the manuscript. All authors have read and agreed to publish this version of the manuscript.

\section*{Competing interests} The authors declare no competing interests.

\end{document}